\documentclass[12pt]{iopart}
 \usepackage{graphicx}
 \usepackage{amssymb}
 \usepackage{color} 
 \usepackage{ulem}
\renewcommand{\vec}[1]{\mathbf{#1}}
\begin{document}
\title[Collimation of a spherical collisionless particles stream]
{Collimation of a spherical collisionless particles stream 
in Kerr space-time}
\author{Kentaro Takami\footnote[1]{Email:
 \mailto{takami@theo.phys.sci.hiroshima-u.ac.jp}}
and Yasufumi Kojima\footnote[2]{Email:
 \mailto{kojima@theo.phys.sci.hiroshima-u.ac.jp}} 
}
\address{Department of Physics,
Hiroshima University,
Higashi-Hiroshima 739-8526, Japan}
%
\begin{abstract}
  We examine the propagation of collisionless particles
emitted from a spherical shell to infinity.
The number distribution at infinity,
calculated as a function of the polar angle,
exhibits a small deviation from uniformity.
The number of particles moving from the polar region 
toward the equatorial plane is slightly larger than that of
particles in the opposite direction,
for an emission radius $ > 4.5M$ in extreme Kerr space-time.
This means that the black hole spin exerts an anti-collimation 
effect on the 
particles stream propagating along the rotation axis.
We also confirm this property in the weak field limit.
The quadrupole moment of the central object
produces a force toward the equatorial plane.
For a smaller emission radius $r<4.5M$,
the absorption of particles into the black hole, the non-uniformity 
and/or the anisotropy of the emission distribution become much more important. 
\end{abstract}
\section{ Introduction }

 Astrophysical relativistic jets are commonly observed in AGNs,
microquasars and gamma ray bursts. The Lorentz factor of the 
outflow motion is $\gamma \approx 10$ in AGNs, 
$\gamma \approx 2-5 $ for microquasars and $\gamma \approx 100$ for
gamma ray bursts. It is considered likely that a central black hole is 
related to the formation of such jets due to the large luminosity 
and short timescale of the jets. However, the 
actual mechanism of such collimated outflows has not yet been resolved, 
although many theories have been proposed. 
One possible mechanism that has been suggested relates to magnetic 
hoop stress. Recently, large-scale MHD simulations of jets 
have been successfully performed by several authors. For example, 
McKinney\cite{mckinney} numerically studied
the hydrodynamical flow emitted from an accretion disk, 
where the jet formation extends to a distance of ten 
thousand times the radius of the central black hole.  
In order for a collimated structure to form, the magnetic pressure
has to be sufficiently large in comparison to the ambient pressure of the 
surrounding medium.

  In this paper, we explore an alternative possibility. 
That is, we investigate a mechanism involving the black hole spin
itself. The strong gravitational field near a black hole also affects
the flow of particles(e.g.,\cite{felice1992,semerak1995}). 
The effect of gravity may be important 
when the emission region 
is concentrated in the vicinity of the central black hole. 
Bi\v{c}\'{a}k et al.\cite{bicak}
discussed this effect by calculating the geodesics of
some particles in Kerr space-time. 
de Felice and Carlotto \cite{felice1997} took into
account the slow loss of angular momentum and energy of the particles
moving along the geodesics.
They derived a certain condition for such dissipative 
collimation and 
discussed its astrophysical relevance \cite{KaDo,deFZ00}. 
Semer{\'a}k\cite{semerak1999} considered 
the effect of a spinning particle in Kerr space-time. 
More recently, Gariel et al.\cite{gariel} discussed the 
collimation of geodesics by using cylindrical coordinates.  
These authors pointed out the possibility of
collimated outflow occurring along the rotation axis 
of a Kerr black hole.
The axis is a unique preferable direction of the flow, and
some trajectories are in fact concentrated along the axis. 
Despite the fact that the possible collimation mechanisms 
in Kerr space-time have been discussed extensively,
it is not clear what fraction of the flow 
(the ratio of the flow along the rotation axis
to the overall emission) is collimated.
For example, if particle emissions are selected with 
orbital parameters suitable for the 
flow along the axis, then the collimated flow is realized.
It is therefore important to evaluate the degree of
the collimation and the results will be
useful for the astrophysical application.

The problem is closely related to the emission process itself.
To tackle this problem, we present a model for studying 
geometrical collimation. 
In our model, billions of particles are isotropically emitted 
from a shell with a constant radius
%
covering a central black hole, and the trajectories are 
numerically calculated. 
We use a test-particle (dust) approximation to calculate
each trajectory.
The collision of particles is ignored in order to evaluate
the geometrical effect in an idealized situation.
The collisional hydrodynamics may work toward local
isotropy, and may  partially hide the effect.
The angular distribution of the particles at the emission radius
is assumed to be uniform, but at infinity some deviation from 
a uniform distribution is produced by Kerr black holes. 
In this way, we study 
to what extent a rotating black hole affects the collimation of
collisionless particles.
Our model is not rigorously realistic,
but is instead an idealized mathematical model which is useful in
evaluating what fraction of the flow is collimated.

  The organization of this paper is as follows. 
In Section \ref{sec:equation}, we summarize our models. 
The numerical distribution of the trajectories of the particles at infinity
is calculated as a function of the polar angle,  
the particles having been emitted with a uniform distribution 
from a 'spherical' shell.   
It is not easy to define a spherically symmetric surface in
non-spherical space-times. Two dimensional spatial surface with a
constant radius is approximately regarded as a sphere.
This definition, which is valid for asymptotically large radius, is
used in this paper.
The numerical results in Boyer-Lindquist coordinate
are given in Section \ref{sec:geometry}.
In order to consider the physical mechanism, the same calculation is
performed in weak gravity. 
Asymptotically Cartesian and mass centered to order N
(ACMC-N) coordinate is used and the results are given 
in Section \ref{sec:weakgravity}.
As the emission radius becomes smaller, a larger number of particles 
which are emitted outwardly are absorbed into the central 
black hole. In this case, the spatial uniformity and
the isotropy of the particle distribution on the emission 
sphere become meaningless.
 In order to demonstrate the absorption effect,
some results are also shown in Section \ref{sec:absorption}.
Finally, a discussion of our results 
is given in Section \ref{sec:discussion}. 
We use units of $c=G=1$ in this paper.

\section{Model and method}
   \label{sec:equation}

The Kerr space-time for a black hole with mass $M$ and angular 
momentum $aM$ is described 
in the Boyer-Lindquist coordinates $(t,r,\theta ,\phi )$ as 
\begin{equation}
ds^{2}=-e^{2\nu }dt^{2}+e^{2\psi }(d\phi -\omega dt)^{2}
+\frac{\rho ^{2}}{\Delta }dr^{2}+\rho ^{2}d\theta ^{2},
  \label{kerrmetric}
\end{equation}
where 
\begin{eqnarray}
&& 
 e^{2\nu }=\frac{\Delta \rho ^{2}}{A}, 
~e^{2\psi }=\frac{ A \sin ^{2}\theta }{\rho ^{2}},
~\omega=\frac{2aMr}{A},
\nonumber \\
&&
\rho ^{2}=r^{2}+a^{2}\cos ^{2}\theta ,
~\Delta =r^{2}+a^{2}-2Mr,
~A=(r^{2}+a^{2})^{2}-a^{2}\Delta \sin ^{2}\theta .
\end{eqnarray}
There are three constants related to the geodesic motion in Kerr
space-time: 
the energy $E$, the $z$ component of the angular momentum 
$L_{z}$ and Carter's constant $\mathcal{Q}$
\cite{misner,Chandrasekhar,frolov}.
The 4-momentum $p_{\mu}$ of a
particle with a rest mass $m$ is therefore written as
\begin{equation}
p_{\mu} = \left( 
-E, ~\pm\frac{R^{1/2}}{\Delta}, ~\pm \Theta ^{1/2}, ~L_z 
\right),
\label{momentum1}
\end{equation}
where 
\begin{equation}
R=(E(r^{2}+a^{2})-L_{z}a)^{2}-\Delta \left[ m^{2}r^{2}
+(L_{z}-aE)^{2}+ \mathcal{Q} \right] ,
\end{equation}
\begin{equation}
\Theta =\mathcal{Q}-\cos ^{2}\theta \left[ 
a^{2}(m^{2}-E^{2})+\frac{L_{z}^{2}}{\sin^{2}\theta} 
\right].
\end{equation}
In order to allow for the emission model of particles to be considered, 
we introduce a
reference frame $(\hat{t},\hat{r},\hat{\theta},\hat{\phi})$ 
measured by zero-angular momentum observers (ZAMO) (e.g.,\cite{frolov}). 
The reference frame is represented
by four orthogonal unit vectors, i.e., one timelike vector 
$\vec{e}_{(0)}$
and three spacelike vectors $\vec{e}_{(n)}(n=1,2,3)$. 
The explicit components are given by 
$e_{(0)}^{\mu }=\left[ e^{-\nu },0,0,\omega e^{-\nu }\right] $, 
$e_{(1)}^{\mu }=\left[ 0,\Delta ^{1/2}/\rho ,0,0\right] $, 
$e_{(2)}^{\mu }=\left[ 0,0,1/\rho ,0\right] $ 
and $e_{(3)}^{\mu }=\left[ 0,0,0,e^{-\psi } \right] $.
The 4-momentum $\vec{p}$ of a particle is expressed by 
\begin{equation}
\vec{p}=\hat{E}\vec{e}_{(0)}+\hat{p}\left(
 \cos \hat{\theta}\vec{e}_{(1)}
+\sin \hat{\theta}\cos \hat{\phi}\vec{e}_{(2)}
+\sin \hat{\theta}\sin 
\hat{\phi}\vec{e}_{(3)}\right) ,
  \label{momentum2}
\end{equation}
where the angles $\hat{\theta}$ and $\hat{\phi}$ are determined 
by the emission direction of the particle. The energy and
the linear momentum satisfy the
special relativistic relation 
$\hat{p}=(\hat{E}^{2}-m^{2})^{1/2}$, 
$\hat{E}=m\hat{\gamma}$, 
where $\hat{\gamma}=\left( 1-\hat{v}^{2}\right) ^{-1/2}$ and 
$\hat{v}$ is the velocity of the test particle.

  From Eqs.(\ref{momentum1}) and (\ref{momentum2}), the constants of motion 
$E$, $L_{z}$, $\mathcal{Q}$ can be expressed 
by quantities measured by a ZAMO,
\begin{eqnarray}
&& E = 
\hat{E}e^{\nu }+\omega L_{z}~,
  \label{constantE} \\
&& L_{z} = 
\hat{p}\sin \hat{\theta}\sin \hat{\phi}e^{\psi},
  \label{constantL}
\\
&&\mathcal{Q} = 
\left( \hat{p}\rho \sin \hat{\theta}\cos \hat{\phi}\right)
^{2}+\cos ^{2}\theta 
\left[ a^{2}(m^{2}-E^{2})
+\frac{L_{z}^{2}}{\sin^{2}\theta} \right].
  \label{constantC}
\end{eqnarray}

We consider a particle moving along the geodesic curve 
without any other forces,
so that $r$-$\theta $ motion can be calculated 
by integrating the differential equation 
\begin{equation}
\frac{d\theta }{dr}=\pm \sqrt{\frac{\Theta }{R}},  
  \label{motion}
\end{equation}
with initial condition 
$(r,\theta )=(r_{\mathrm{e}},\theta _{\mathrm{e}})$.
The numerical integration method is a fourth-order Runge-Kutta method
with an adaptive step size, and the numerical errors are adjusted to be
small enough.
In order to check the validity of our numerical integration,
we calculated the same problem as Bi\v{c}\'{a}k et al.\cite{bicak}
and obtained the same results.
  We consider the particle distribution, 
$f(\cos \theta )$, at infinity 
$r_{\infty }$, when a large number of test particles, 
$N_{\mathrm{tot}}$,
are emitted from a shell of radius $r_{\mathrm{e}}$.
The particle distribution at $r_{\mathrm{e}}$ is 
assumed to be uniform, 
that is, the number of particles $dN$ emitted 
from $\theta _{\mathrm{e}}$ to 
$\theta _{\mathrm{e}}+d\theta _{\mathrm{e}}$ 
is $dN=(N_{\mathrm{tot}}/4\pi )(2\pi \sin \theta _{
\mathrm{e}}d\theta _{\mathrm{e}})$ for 
$0\leqq \theta _{\mathrm{e}}\leqq \pi $. 
As for the emission model at $(r_{\mathrm{e}},\theta _{\mathrm{e}})$, 
we also assume that the particles have the same energy $\hat{E}$, 
and that the emission direction is isotropic with respect 
to the angles $\hat{\theta}$
and $\hat{\phi}$. 
The angle $\hat{\theta}$ might be limited to 
$0\leqq \hat{\theta}\leqq \pi /2$, since most of the particles 
satisfying the emission angle $\pi /2\leqq \hat{\theta}\leqq \pi $ 
are absorbed by the black hole and can therefore be ignored from the beginning. 
In the numerical calculation, the directions 
$\hat{\theta}$ and $\hat{\phi}$ are randomly chosen 
at $r_{\mathrm{e}}$, and the constants of motion can be
calculated by 
Eqs.(\ref{constantE})-(\ref{constantC}).
The asymptotic value $\theta $ of each particle is numerically 
evaluated by solving Eq.(\ref{motion}) to $r_{\infty }$, 
where $r_{\infty }$ is a
large number satisfying $r_{\infty }\gg M$, which is set to `infinity' 
in the numerical calculation. By calculating $N_{\mathrm{tot}}$ 
trajectories, we obtain the number distribution $f_*(\cos \theta )$.

In order to study the geometrical effects, 
we are not interested in a fine-grained distribution $f_*$
but in a coarse-grained distribution, i.e., an averaged distribution 
on a larger angular scale.
  In the numerical calculations, we divide 
$\cos \theta (0\leqq \cos \theta \leqq 1)$ into $N_{0}$ segments 
with a small interval $d\cos \theta$, and
count the number of the trajectories which arrive at the region 
between $\cos\theta $ and $\cos \theta +d\cos \theta $. 
The number is approximated by 
$f(\cos \theta )d\cos \theta $, and the function $f(\cos \theta )$ 
is numerically constructed. 

  In our model, the isotropy of the emission direction
and the spatial uniformity are assigned by a ZAMO at 
a constant radius in Boyer-Lindquist coordinates.
If the emission is uniform 
with respect to the proper surface element at
$r_{\mathrm{e}}$, then  
the number of particles $dN$ should be modified as
$ dN=g(N_{\mathrm{tot}}/4\pi )
(2\pi \sin \theta _{\mathrm{e}}d\theta _{\mathrm{e}})$ 
for $0\leqq \theta _{\mathrm{e}}\leqq \pi $.
The correction factor $g$ is estimated as 
$g=2
A^{1/2}/(\int_{0}^{\pi}A^{1/2}\sin \theta d\theta )$
$\approx1+a^2(r_{\mathrm{e}} -2M)(3 \cos ^2\theta -1)/(6r_{\mathrm{e}}^{3})$,
which gives a small variation of 
$5 \times 10^{-3}$ for $r_{\mathrm{e}} = 4.5M$. 
The correction $|g-1|$ seems to increase as $r_e$ decreases.
We have numerically evaluated the factor and found that
there is a maximum value at a certain radius $r_e$. For example,
in extreme Kerr space-time $a=M$, we found that 
$|g-1|< 1.5 \times 10^{-2}$, where the maximum occurs at
$r_e \sim 2.4M$ and $\theta =0, \pi$.
We actually include this correction factor into
the numerical calculations in Section 3 and 5,
and the obtained results are not significantly different, since
other physical effects are much larger in the magnitude. 

%
%
%

\section{Effect of geometry on propagation}
\label{sec:geometry}

In principle, the number distribution function
 $f(\cos \theta )$ ranges from
0 to $N_{\mathrm{tot}}$ and clearly depends on $N_{\mathrm{tot}}$.
 Therefore, this
property is not suitable when comparing, for example, 
numerical results obtained using
different total numbers of particles. 
The deviation from uniform distribution is
more important. We therefore use the following normalized distribution 
$F(\cos \theta )$ defined by 
\begin{equation}
F(\cos \theta )=\frac{f(\cos \theta )-\langle f\rangle }{\langle f\rangle }
\times 100[\%],
\label{norm}
\end{equation}
where 
\begin{equation}
\langle f\rangle =
\frac{\int_{0}^{1}f(\cos \theta )d(\cos \theta )}
{\int_{0}^{1}d(\cos \theta )}.
\end{equation}
This normalized function becomes $F=0$ when $f$ does not depend on 
$\cos \theta$, i.e., in the case of a spherically symmetrical distribution.

\begin{figure}[tbp]
\begin{center}
\includegraphics[scale=0.5]{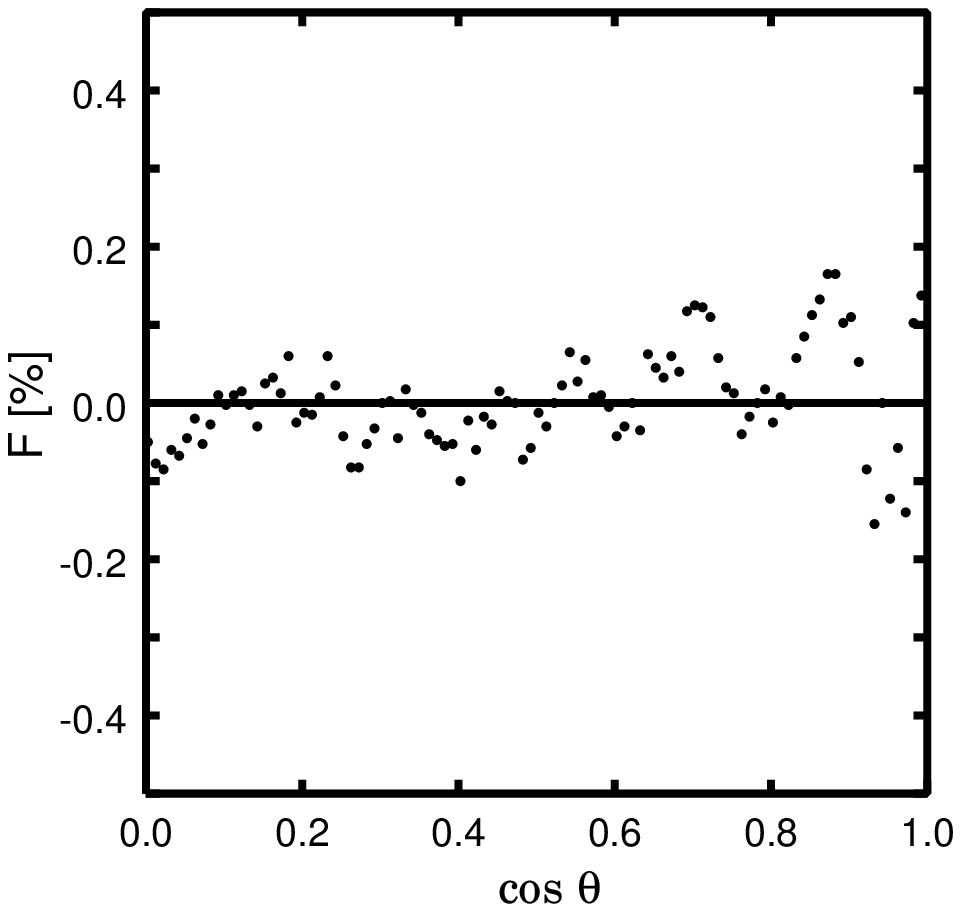}
\includegraphics[scale=0.5]{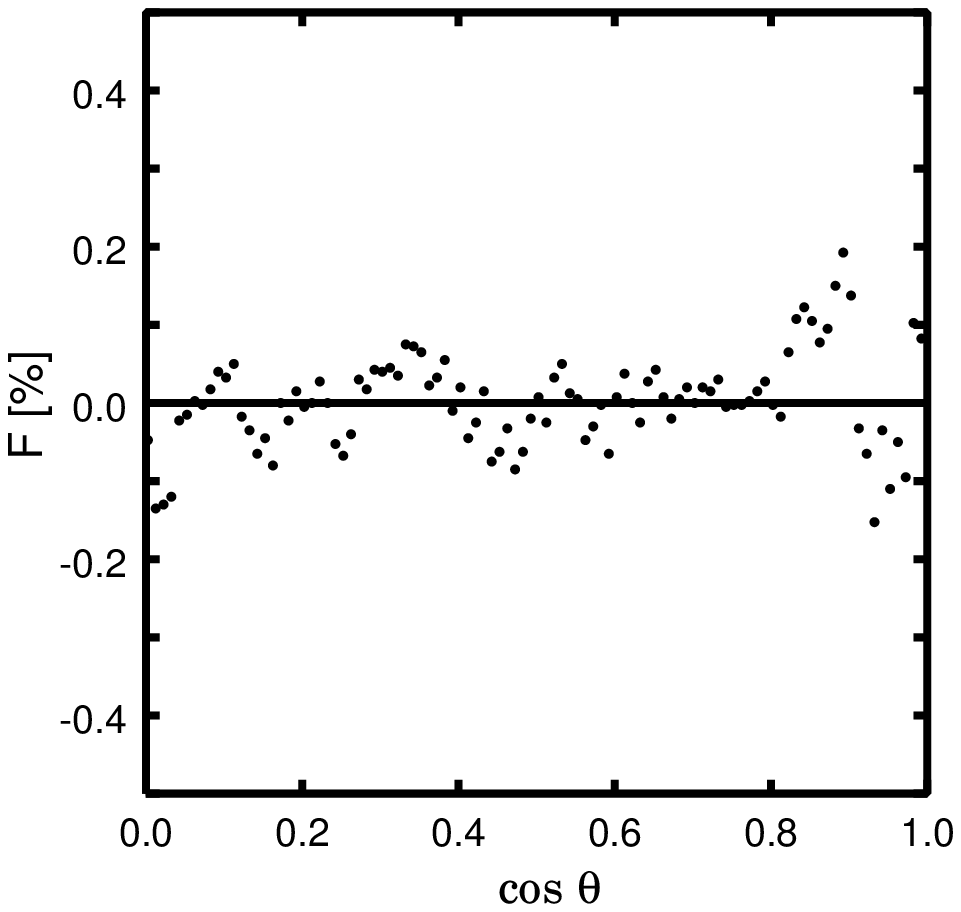}
\end{center}
\caption{Deviation from uniform distribution as a function of 
$\cos \theta $. 
The left panel shows the results obtained for a flat space-time, 
and the right panel shows the results obtained for a Schwarzschild space-time.}
\label{Fig01}
\end{figure}

First, we present the two test problems. The distributions are calculated
for a flat space-time and a Schwarzschild space-time.  
We expect $F=0$,
for $0\leqq \cos \theta \leqq 1$ since both space-times have 
spherical symmetry. 
We adopt $N_{\mathrm{tot}}=2\times 10^{9}$, $r_{\mathrm{e}}=4.5M$, 
$r_{\infty }=500M$ and $\hat{v}=0.9~(\hat{\gamma}\sim 2)$ 
in the numerical calculations. The results are shown 
in Figure~\ref{Fig01}.
There are small irregular deviations from $0$ in
the number distributions $F$. These small deviations are $\sim0.2\%$ at most, 
and arise from the finite numbers $N_{\mathrm{tot}}$ and $N_{0}$. 
In constructing $F$, we divide the angle into $N_{0}(=100)$
segments. If $N_{\mathrm{tot}}(=2\times 10^{9})$ trajectories are randomly
distributed, then the mean value is $N_{\mathrm{tot}}/N_{0}$, and the
deviation from the average is 
$(N_{0}/N_{\mathrm{tot}})^{1/2}\approx 2\times 10^{-4}$. 
Additionally, randomly chosen emission angles, 
$\hat{\theta}$ and $\hat{\phi}$, are involved in the numerical
calculation, so that the deviation increases by a certain factor. 
We numerically checked that the statistical error is roughly
proportional to $N_{\mathrm{tot}}^{-1/2}.$ 
The error therefore becomes $0$ 
in the limit of $N_{\mathrm{tot}}\rightarrow \infty $. 
Fortunately, this `noise' level is sufficiently low to allow the effect 
of black hole spin to be examined, as discussed below.

\begin{figure}[tbp]
\begin{minipage}{0.45\linewidth}
%
\includegraphics[scale=0.5]{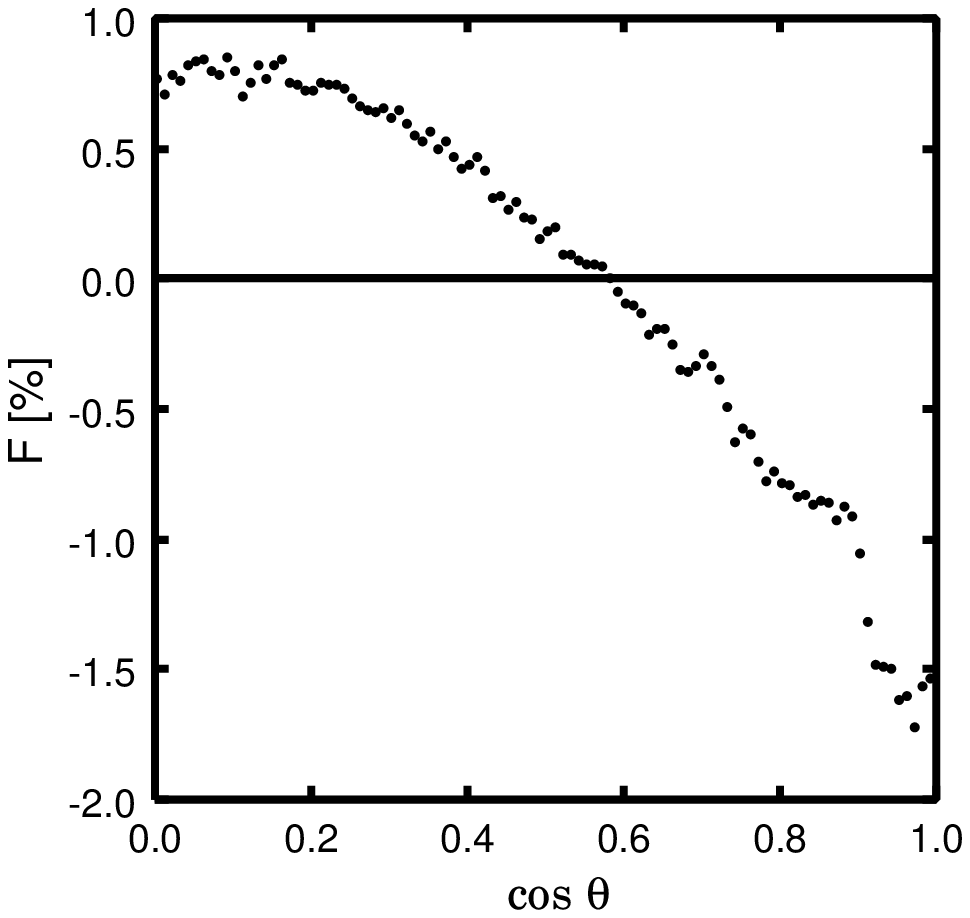}
\caption{The same as Figure~\ref{Fig01},
 but for the extreme Kerr space-time.}
\label{Fig02}
\end{minipage}
\hspace{5mm}
\begin{minipage}{0.45\linewidth}
\includegraphics[scale=0.50]{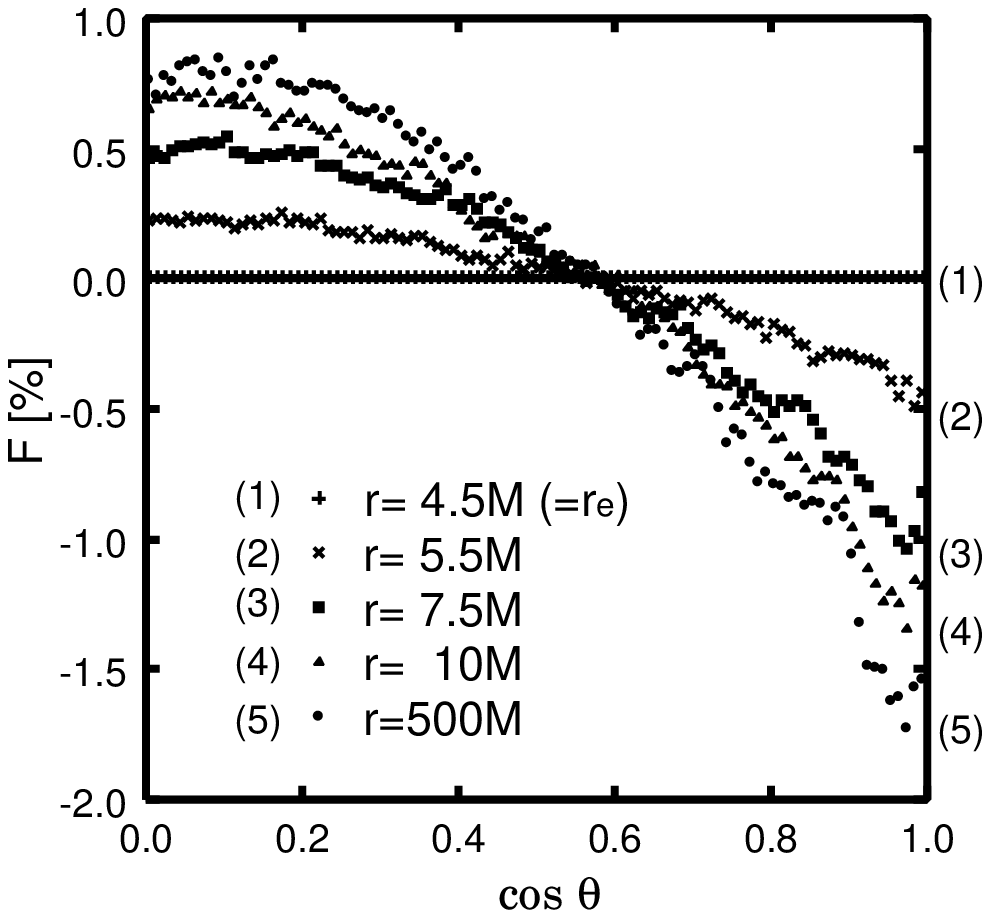}
\caption{Evolution of normalized number distribution for
different values of the radial coordinate $r$.}
\label{Fig03}
\end{minipage}
\end{figure}

 Next, we calculate the number distribution $F$ for 
the extreme Kerr space-time. We adopt the same parameters 
as those used in the test calculations.
The results are shown in Figure~\ref{Fig02}. 
The number distribution $F$ exhibits a larger deviation, where 
there is a deficit of $F\sim -1.7\%$ near the polar regions and
an excess of $F\sim 0.8\%$ near the equatorial region. 
These values are obviously larger than the statistical 
error of $\sim 0.2\%$ in test calculations. 
The results, however, are opposite to our expectation that 
the spin of the black hole would cause the particles 
to be collimated along the rotation axis.

 Figure~\ref{Fig03} shows how the number distribution $F$ changes 
with the propagation distance. 
The distributions for some selected radii are calculated 
from an initially uniform distribution.
It can easily be seen that 
the deviation from the average gradually increases
with $r$ from the emitting radius $r_{\mathrm{e}}=4.5M$ to 
$r_{\infty }=500M$. 
This increase almost stops for $r>10M$, where the effect of the Kerr
space-time becomes very small $(a/r)^{2}\sim 10^{-2}$. 
This result indicates that the particles tend
to fall toward the equatorial plane due to the Kerr black hole. 
This feature has been partially discussed 
by Bi\v{c}\'{a}k et al.\cite{bicak}.

In order to express the above geometrical deviation
in a simple way, we calculate the function $F$ 
for various values of the Kerr parameter $a$, the emitting 
radius $r_{\mathrm{e}}$ and the emitting velocity $\hat{\gamma}$.
It is found that the following empirical relation is well fitted 
to the numerical results of approximately 20 models
in the ranges 
$0\leq a\leq M$, $4.5M\leq r_{\mathrm{e}}\leq 10M$ and 
$2\leq \hat{\gamma}\leq 10^{3}$,
\begin{equation}
F(\cos \theta )\sim -9.2\times 10
\left( \frac{a}{M}\right) ^{2}\left( 
\frac{r_{\mathrm{e}}}{M}\right) ^{-2.5}
\left( 1+\frac{1}{\hat{\gamma}^{2}}
\right) \left( \cos ^{2}\theta -\frac{1}{3} \right),
 \label{empiricaleq}
\end{equation}
where the angular part is the Legendre polynomial of $l=2$.
 From this relation, the most important factor 
affecting the geometrical deviation is the emission 
radius $r_{\mathrm{e}}$. 
Our calculation in this section is
limited to $r_{\mathrm{e}}\geq 4.5M$, 
but the deviation can be increased further for particles 
emitted in the vicinity of the black hole. 
As shown in Section \ref{sec:absorption}, 
however, the extrapolation of Eq.(\ref{empiricaleq}) 
for smaller values of $r_{\mathrm{e}}$ is not particularly 
meaningful, since the fraction of absorbed trajectories 
becomes more important.

\section{Weak gravity with quadrupole moment}
\label{sec:weakgravity}

\begin{figure}[tbp]
\centering
\includegraphics[scale=0.5]{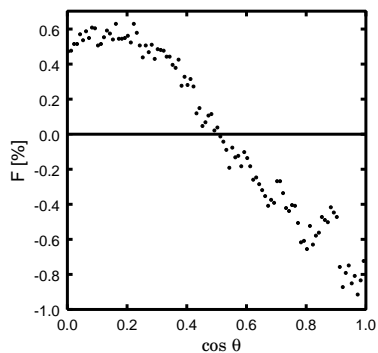}
\caption{
Deviation from uniform distribution at infinity for $a=M$ 
in the ACMC-2 coordinate, 
$r^{'}_{\mathrm{e}}=4.5M$ and $\hat{v}=0.9~(\hat{\gamma}\sim 2)$. 
}
\label{fig:figACMC}
\end{figure}

In order to examine the deviation from spherical symmetry, we consider 
the weak gravity limit of Kerr space-time.
We introduce asymptotically Cartesian and mass centered to order N
(ACMC-N) coordinate system$(t',r',\theta',\phi')$ 
\cite{thorne1980}.
In the ACMC-2 coordinate system, which is the lowest order containing
non-spherical terms, the Kerr metric is written as follows:
\begin{eqnarray}
g_{0'0'} &=&
 -1+\frac{2M}{r'}-\frac{3Ma^2\cos^2\theta'}{{r'}^3}  
~,\label{eq:ACMCg00} \\
g_{0'3'} &=& r' \sin\theta' 
\left( 
-\frac{2Ma\sin\theta'}{{r'}^2} 
+\frac{5Ma^3\sin\theta'\cos^2\theta'}{{r'}^4} 
\right)~, \\
g_{1'1'} &=& 1+\frac{2M}{r'}+\frac{4M^2-a^2}{{r'}^2}
+\frac{8M^3-4Ma^2-Ma^2\cos^2\theta'}{{r'}^3}
~, \\
g_{2'2'} &=& {r'}^2 
\left( 1+\frac{a^2}{{r'}^2} 
\right)~, \\
g_{3'3'} &=& {r'}^2 \sin^2 \theta'
\left(
1+\frac{a^2}{{r'}^2}+\frac{2Ma^2\sin^2\theta'}{{r'}^3}
\right)~, \\
g_{1'2'} &=& r' 
\left( -\frac{2Ma^2\cos\theta'\sin\theta'}{{r'}^3} 
\right), \label{eq:ACMCg12}
\end{eqnarray}
where the relations to the Boyer-Lindquist coordinates 
$(t,r,\theta ,\phi )$ are
\begin{eqnarray}
&& r=r' +\frac{a^2\cos^2\theta'}{2r'}, ~~~
\theta = \theta' -\frac{a^2\cos\theta'\sin\theta'}{2{r'}^2}, 
\nonumber \\
&& \phi=\phi',~~~ t=t'. 
\end{eqnarray}
One of the important non-spherical effects is the quadrupole field in the
Newtonian level, $g_{0'0'}=-(1+2\Phi_N)$, where
\begin{equation}
\Phi_N=-\frac{M}{r'}+\frac{3Ma^2 \cos^2\theta'}{2r'^3}~.
\end{equation}
A particle in this potential field receives an acceleration in the 
$\theta'$-direction as 
\begin{equation}
\alpha_{\theta'} = \frac{3Ma^2\sin 2\theta'}{2{r'}^4}~.
\label{eq:force}
\end{equation}
This is positive in the upper half plane($0<\theta'<\pi/2$),
while negative for $\pi/2<\theta'<\pi$.
The quadrupole field therefore bends the orbit of a particle toward 
the direction of equatorial plane.
The quadrupole feature of gravity should be one of the mechanisms
for the anti-collimation shown in Section \ref{sec:geometry}.
In the following, we examine numerically the magnitude of the
deviation in the ensemble of particles which is caused by the quadrupole
moment. 
We calculate the particle distribution $F(\cos \theta )$ 
using Eqs.(\ref{eq:ACMCg00})-(\ref{eq:ACMCg12}) and geodesic equations, 
in the same way as done in Section \ref{sec:geometry}.
In this space-time, we solve the second-order 
differential equation for the geodesic
without using Carter's constant ${\cal Q}$.
The results for 
$a=M, r'_{\mathrm{e}}=4.5M$ and $\hat{v}=0.9~(\hat{\gamma}\sim 2)$
are shown in Figure \ref{fig:figACMC}.
The number distribution $F$ has a deficit near the polar region and an
excess near the equatorial region.
This behavior is the same as Figure \ref{Fig02}.
The concentration of particles toward the equatorial plane can 
partially be explained by the mass quadrupole moment of the central object.
The deviation of Figure \ref{Fig02} is roughly twice as large as that of
Figure \ref{fig:figACMC}. It is impossible to exactly
compare results in different space-times exactly. 
The emission models are different.
For example, 
the constant radius $r'_{\mathrm{e}}=4.5M$ in ACMC-2 coordinate does not
agree with the radius $r_{\mathrm{e}}=4.5M$ in the Boyer-Lindquist
coordinate. 
The correspondence is approximately valid. 
There is a common feature of particle
concentration
toward the equatorial plane, although 
exact comparison between two different coordinates is impossible,
and hence there is still a problem in the magnitude. 
In the weak field limit, the
feature is explained by a mass quadrupole term. The feature still appears
in much strong field, as calculated in Section \ref{sec:geometry}.

\section{Absorption effect}
\label{sec:absorption}

\begin{figure}[tbp]
\centering
\includegraphics[scale=0.4]{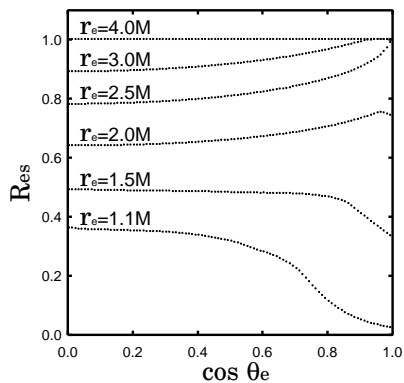}
\caption{
Escape fraction for various
emission radii $ r_\mathrm{e}$
in the extreme Kerr space-time.
The escape fraction is the ratio of the number of particles 
which have escaped to infinity
to the total number of particles emitted 
from $( r_\mathrm{e},  \theta_\mathrm{e})$. }
%
\label{Fig04}
\end{figure}

  The number of orbits in which particles are absorbed into the black hole
increases as the distance between the emitting surface 
and the black hole decreases.
Figure~\ref{Fig04} 
shows the escape fraction $ R_{\mathrm es}$ 
as a function of $ \cos \theta_\mathrm{e}$
in the extreme Kerr space-time.
The escape fraction is the ratio of the number of escaped particles
to the total number of particles emitted 
from $( r_\mathrm{e},  \theta_\mathrm{e})$.
The initial velocity of the particles is chosen as 
$\hat{v}=0.99999~(\hat{\gamma}\sim 224)$.
For a large emission radius, the escape fraction is almost unity, 
which indicates that in-falling orbits to the black hole can be ignored.
As the radius $ r_\mathrm{e}$ decreases,
the escape fraction significantly decreases. 
Figure~\ref{Fig04} shows that
the distribution of particles escaping to infinity or being absorbed into the black hole 
is not uniform, showing a slight
dependence on the angle $ \theta_\mathrm{e}$.
Such dependence indicates that the emission distribution can be regarded as 
being non-uniform.
This property is important since 
the non-uniformity at the emission surface 
naturally leads to the non-uniform outflow at infinity.
The trajectories of particles emitted at an angle $ \theta_\mathrm{e} $
do not necessarily have the same
angle $  \theta =  \theta_\mathrm{e} $
at infinity, as the correspondence 
between the polar angles is somewhat different 
at the time of emission and at the time of observation 
due to its modification during the propagation of the particles.
It is seldom the case that a non-uniform distribution becomes 
a uniform distribution at infinity,
and therefore the existence of non-uniformity at emission
might be important.

\begin{figure}[tbp]
\centering
\includegraphics[scale=0.5]{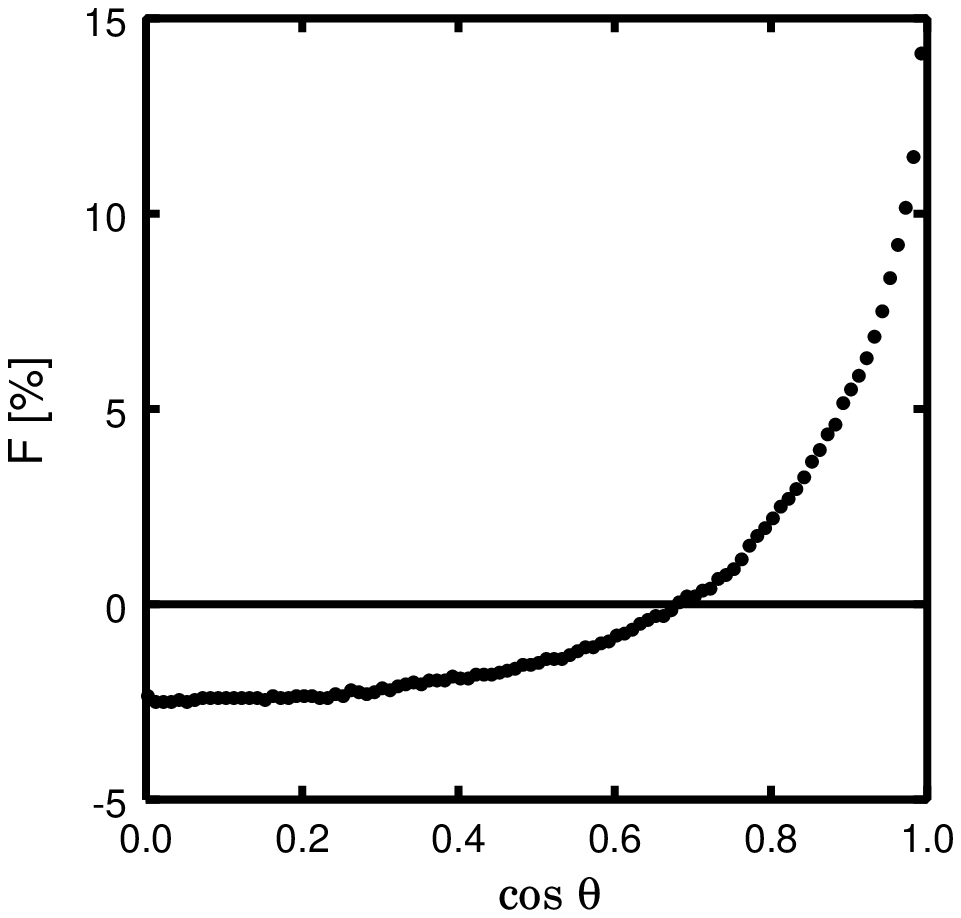}
\includegraphics[scale=0.5]{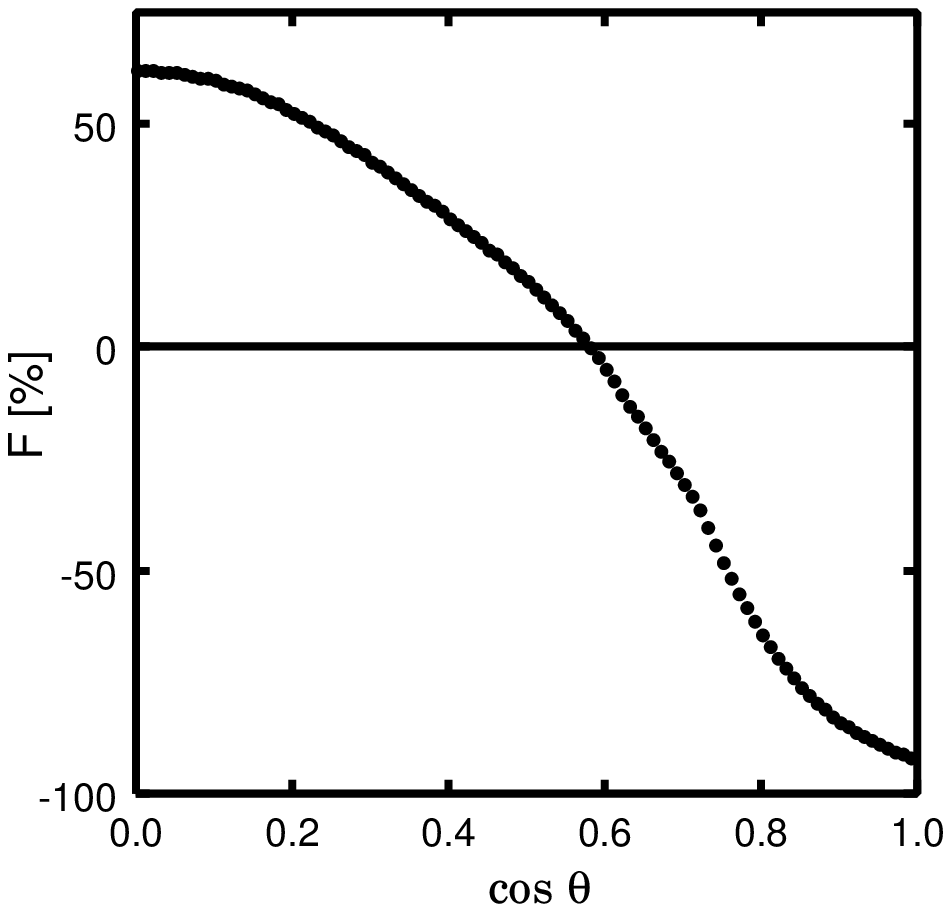}
\caption{
Deviation from uniform distribution at infinity.
The left panel shows the results for emission radius
$r_{\mathrm{e}} = 2.5M$,
and the right for $r_{\mathrm{e}} = 1.1M$.
}
\label{Fig05}
\end{figure}
%

The results shown in Figure~\ref{Fig05} 
illustrate the effect of the orbits in which particles are absorbed
in the extreme Kerr black hole. 
The left and right panels show the results obtained for  
an emission radius $r_{\mathrm{e}} = 2.5M$
and $r_{\mathrm{e}} = 1.1M$, respectively.
The initial velocity is 
$\hat{v}=0.99999~(\hat{\gamma}\sim 224)$,
and the other
parameters are the same as those used in the test calculations.
Regarding emission for $r_{\mathrm{e}} = 2.5M$,  
there is an excess of $F\sim 14\%$ at $ \cos \theta =1$ and
a deficit of $F\sim -2\%$ at $ \cos \theta =0$.
An interesting feature is that 
the excess appears near the polar regions, which
is opposite to the result obtained 
for $r_{\mathrm{e}} \geq 4.5M$,
as considered in Section \ref{sec:geometry}.
Furthermore, the magnitude is much larger than 
the geometrical effect, which is estimated by
extrapolating the empirical formula (\ref{empiricaleq}),
and is $F \sim -6\%$ at the largest deviation point
$ \cos \theta =1$.
Thus, it is concluded that  the number distribution strongly depends on other
factors.
As shown in Figure~\ref{Fig04},
the escape fraction for $r_{\mathrm{e}} = 2.5M$ 
varies with a typical amplitude $ \pm 10\%$
around the average value.
This effect overwhelms the modifications which occur during the propagation, 
although the excess at the polar regions is expected 
to be suppressed to a certain extent if the extrapolation still holds.
The right panel of Figure~\ref{Fig05}
shows the results obtained for a smaller emission 
radius $r_{\mathrm{e}} = 1.1M$.
Note that the outer boundary of the ergoregion is given by 
$ r=M(1+\sin \theta) $ in the extreme Kerr case,
where the emission sphere is located 
within the ergoregion, except for the polar regions. 
The excess around the polar regions  
transforms into a great deficit of $-90\%$, and
the deviation from the average also becomes much larger.
The origin of this feature can mainly be explained by 
the non-uniformity at the emission surface $r_{\mathrm{e}}$.
%

\section{Discussion}
  \label{sec:discussion}

%
  A strong gravitational field affects the emission and motion of 
particles in several ways.
For example, a central black hole plays an important role
in forming the surrounding accretion disk, 
and hence determines the distribution of the particles which are emitted
into phase space, 
i.e., the distributions of position and velocity.
Strong gravitational fields also affect the propagation of particles.
In this paper, we have concentrated on the latter in 
a Kerr space-time by calculating the trajectories of particles emitted 
from a `spherical' shell.
Our concern is not the orbit of a single particle, but 
the behavior of an assembly of particles.
That is, the deviation from `spherical symmetry' which occurs
during the propagation of the particles which are initially
uniformly distributed on the emission shell and are  
emitted in isotropic directions. 
 We were able to extract the geometrical effect
under an ideal situation for the emission at 
$r_\mathrm{e}\gtrsim 4.5M $.
In this case, 
the particles are anti-collimated on average,
i.e., they spread in a direction perpendicular to
the rotation axis in Kerr space-time. 
This feature can be seen even in weak gravitational field with mass
quadrupole moment, which partially accounts for the physical mechanism.
The magnitude depends on the Kerr parameter
and the emitting radius $r_\mathrm{e} $, 
and 
the deviation from uniform distribution
amounts to a few percent in the case of
$r_\mathrm{e} = 4.5M $.
With a further decrease in the radius, the geometrical effect 
increases, but the other effects, 
such as the absorption of particles in certain orbits,
become important, 
and therefore our definition of collimation/anti-collimation
becomes ambiguous for $r_\mathrm{e} \lesssim 4.5M $.

 This feature of anti-collimation along the rotation axis
does not agree with the results of previous works
at first sight. 
The main difference arises from the emission models used.
Bi\v{c}\'{a}k et al.\cite{bicak} derived orbital parameters of which a
particle emitted in the radial direction is collimated toward the 
rotation axis. The condition 
satisfies  $\mathcal{Q} <0$, which corresponds to the $\theta$ motion
confined in the range $ 0 < \theta < \pi/2$
or $ \pi/2   < \theta < \pi $ \cite{Chandrasekhar}.
We checked the condition in our numerical simulation, 
and found that the fraction of trajectories satisfying 
it is typically $\sim 0.1$\%. 
The isotropic emission is assumed in our model, so that
most trajectories are irrelevant to the collimation
condition derived by Bi\v{c}\'{a}k et al.\cite{bicak}.
Williams\cite{Williams} calculated
the production and subsequent propagation of $e^-$ $e^+$ pairs
in the Penrose process within the ergosphere by
using a Monte Carlo simulation.
The results show that escaping trajectories
are collimated and that helical trajectories exist along the rotation axis.
The initial emission within the ergosphere
is highly constrained in the Penrose process\cite{Bardeen},
since one of the particles in the pair should be in 
a negative energy state.
The emission model is therefore different from ours. 
Recently, Bini et al.\cite{bini}
considered the outflow along the rotation axis, 
and showed that an initially spherical assembly of particles 
becomes elongated along the rotation axis 
during the propagation of the particles.
In their model, the initial direction of the velocity is fixed along
the rotation axis, which is again different from our model.

It is not easy to summarize the effects of a strong gravitational
field on the collimation of emitted particles
in a Kerr space-time while taking into account the results 
obtained in other works.
In our model, the effect which occurs during the propagation
has been clarified as a result of simplifying the emission model.
We found the unanticipated result, i.e., no collimation along the spin
axis, but rather spread toward equator.
This anti-collimation effect along the rotation axis is mainly explained by
mass quadrupole moment of the Kerr space-time. 
The geometrical modification integrated from
the emission point to infinity is 
$ \sim (a/M)^2 ( 6M/r_\mathrm{e})^{2.5}$, which is not so large.
We conclude that in order for the collimated outflow to form, 
the emission model should provide a much larger deviation
in the emitting direction and in the spatial distribution,
where the Kerr space-time has a considerable effect.  

\ack
  Numerical computation in this work was carried out using the 
PC Cluster System of the Information Media Center 
in Hiroshima University. This work was supported in part
by a Grant-in-Aid for Scientific Research
from the Japan Society for the Promotion of Science (JSPS) (KT).

 \section*{References}

   \end{document}